\documentclass[twocolumn,conference]{IEEEtran}
\usepackage[latin9]{inputenc}
\usepackage{units}
\usepackage{amsmath}
\usepackage{stmaryrd}
\usepackage{stackrel}
\usepackage{graphicx}
\usepackage[unicode=true,
 bookmarks=true,bookmarksnumbered=true,bookmarksopen=true,bookmarksopenlevel=1,
 breaklinks=false,pdfborder={0 0 0},pdfborderstyle={},backref=false,colorlinks=false]
 {hyperref}
\hypersetup{pdftitle={Your Title},
 pdfauthor={Your Name},
 pdfpagelayout=OneColumn, pdfnewwindow=true, pdfstartview=XYZ, plainpages=false}

\makeatletter
\usepackage[caption=false,font=footnotesize]{subfig}
\usepackage{cite}

\makeatother

\begin{document}
\title{Mobility-aware Beam Steering in Metasurface-based Programmable Wireless
Environments}
\author{Christos Liaskos\IEEEauthorrefmark{1}\IEEEauthorrefmark{4}, Shuai
Nie\IEEEauthorrefmark{3}, Ageliki Tsioliaridou\IEEEauthorrefmark{1},
Andreas Pitsillides\IEEEauthorrefmark{2}, Sotiris Ioannidis\IEEEauthorrefmark{1},
and Ian Akyildiz\IEEEauthorrefmark{2}\IEEEauthorrefmark{3}\\
{\small{}\IEEEauthorrefmark{1}Foundation for Research and Technology
- Hellas (FORTH), emails: \{cliaskos,atsiolia,sotiris\}@ics.forth.gr}\\
{\small{}\IEEEauthorrefmark{4}University of Ioannina, Computer Science
and Engineering Department, Greece}\\
{\small{}\IEEEauthorrefmark{2}University of Cyprus, Computer Science
Department, email: Andreas.Pitsillides@ucy.ac.cy}\\
{\small{}\IEEEauthorrefmark{3}Georgia Institute of Technology, School
of Electrical and Computer Engineering, emails: \{shuainie,ian\}@ece.gatech.edu}}
\maketitle
\begin{abstract}
Programmable wireless environments (PWEs) utilize electromagnetic
metasurfaces to transform wireless propagation into a software-controlled
resource. In this work we study the effects of user device mobility
on the efficiency of PWEs. An analytical model is proposed, which
describes the potential misalignment between user-emitted waves and
the active PWE configuration, and can constitute the basis for studying
queuing problems in PWEs. Subsequently, a novel, beam steering approach
is proposed which can effectively mitigate the misalignment effects.
Ray-tracing-based simulations evaluate the proposed scheme.
\end{abstract}


\IEEEpeerreviewmaketitle{}

\section{Introduction\label{sec:Introduction}}

Software-defined Metasurfaces (SDMs) have received considerable attention,
due to their efficiency in exerting control over electromagnetic (EM)
waves during their propagation from a transmitter to a receiver~\cite{CACM.2018}.
SDMs allow for precise control over the direction, polarization, amplitude
and phase of waves impinging over them, in a frequency and encoding-selective
manner~\cite{MSSurveyAllFunctionsAndTypes,zhang2018space,Liu.2019g}.
Moreover, metasurface hypervisors called HyperSurfaces (HSFs) have
been proposed, which allow for chaining and exerting many wave manipulation
functionalities at once via well-defined application programming interfaces~\cite{Liaskos.TNET.2019}.

These exquisite capabilities have recently enabled the PWE concept~\cite{CACM.2018}.
PWEs are created by applying HSF coating over large planar objects,
such as ceilings and walls in floorplans. A central server orchestrates
the wave manipulation functions per HSF unit, thereby customizing
the wireless propagation \emph{as an app}, tailoring to the locations
and needs of the users. Mitigating the NLOS and distance problems
in mm-wave communications~\cite{Mehrotra.2019,Alexandropoulos.2019},
negating fading phenomena and Doppler Effects, advanced physical-layer
security, as well as enabling long-distance wireless power transfer~\cite{Liaskos.TNET.2019}
are indicative PWE capabilities.

\begin{figure}[htbp]
\begin{centering}
\textsf{\includegraphics[viewport=0bp 70bp 450bp 540bp,clip,width=1\columnwidth]{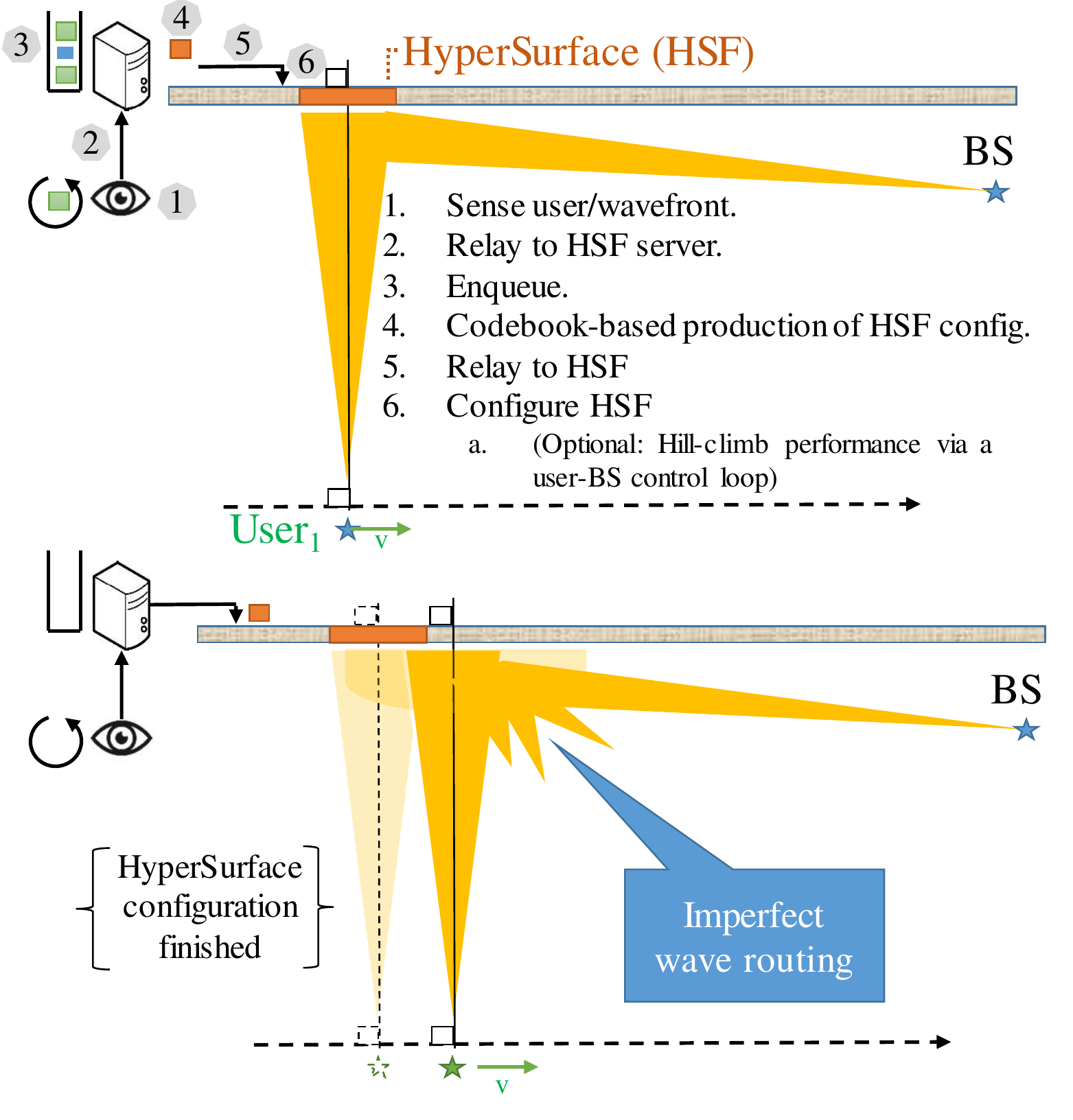}}
\par\end{centering}
\caption{\label{fig:setup}The considered PWE adaptation cycle, comprising
a server that senses the wave emissions from a user and re-configures
the HSF accordingly (top). The inserted latency can give rise to misalignment
between the user emissions and the HSF configuration (bottom). }
\end{figure}
In order to optimally configure a PWE, the central server needs to
sense the wavefronts emitted from the user devices. This task can
be handled by the HSFs themselves using embedded sensors~\cite{absense.2019},
or in a sensor-less HSF-synergistic fashion fashion based on compressed
sensing principles~\cite{Liaskos.2019b}. In mobility scenarios,
however, as shown in Fig.~\ref{fig:setup}, a delay in completing
the sense/process/configure cycle may result into misaligned user
emissions and HSF configurations, yielding imperfect wave routing.

The contribution of the present study is twofold: first, we model
the delay of the HSF sense/reconfigure cycle, accompanied by an analytical
approach to study the misaligned beam steering effects. Subsequently,
we propose a beam steering configuration that can effectively account
for such misalignment, exploiting the high-precision wave manipulation
capabilities of SDMs. Ray-tracing-based simulations evaluate the effectiveness
of the proposed scheme, also covering cases where some degree of the
misalignment phenomenon can be accounted for via user position prediction.

\section{System Model\label{sec:System-Model}}

\begin{figure}
\begin{centering}
\textsf{\includegraphics[width=1\columnwidth]{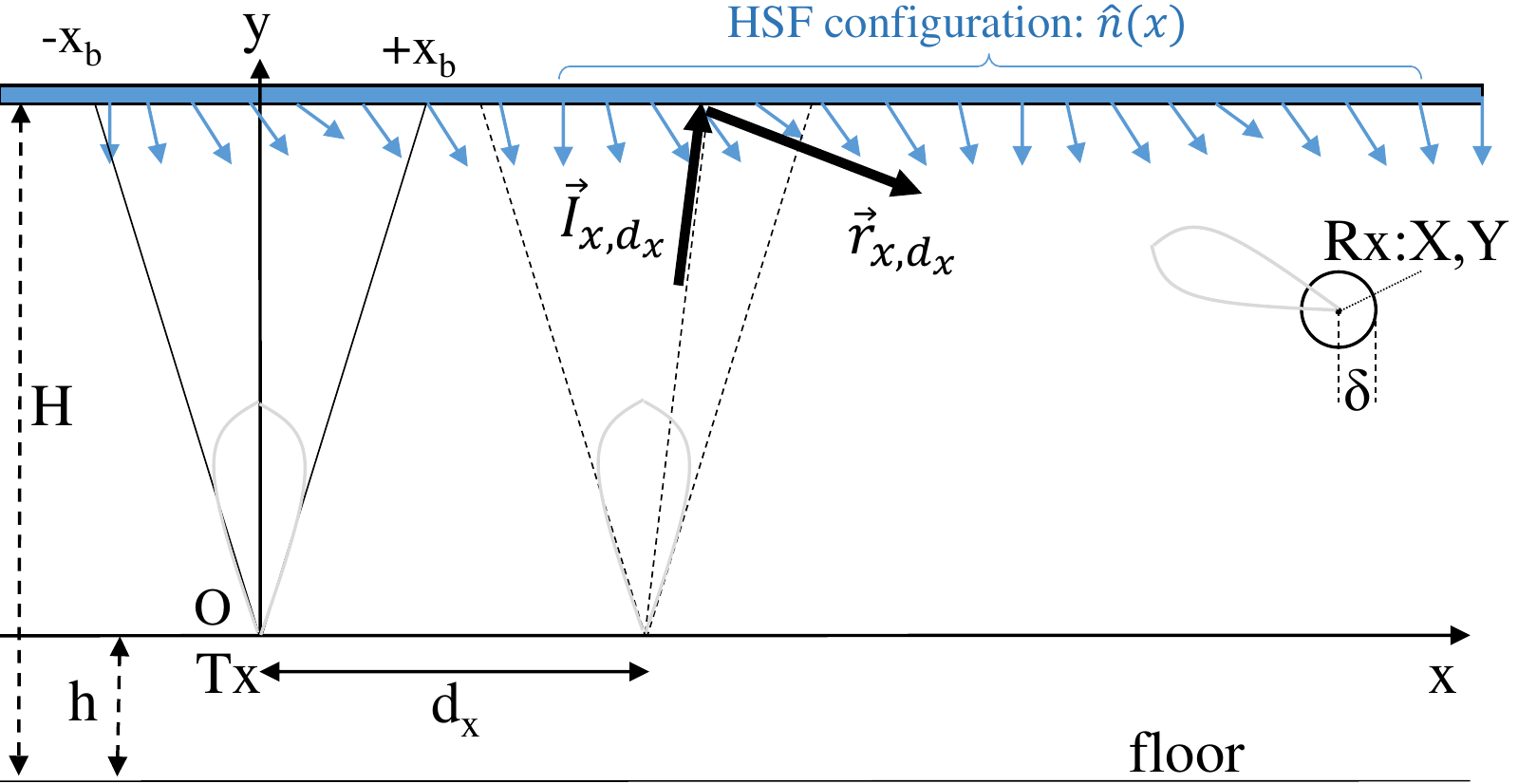}}
\par\end{centering}
\caption{\label{fig:notation}Illustration of the basic notation employed.}
\end{figure}
Consider the setup of Fig.~\ref{fig:setup} comprising a corridor
where the ceiling is HSF-coated, a user moving through it, and a PWE
server controlling the HSF configuration adaptively. Ceilings are
promising areas to apply HSF coating since they: i) constitute large
areas within any floorplan, ii) they are commonly unused, iii) they
provide easy access to power supply, and iv) facilitate contact with
waves created by user devices such as smartphones: Assuming that the
gyroscope of a user device can detect the upwards direction, the user
device can subsequently beamform upwards always, meeting the HSF-coated
ceiling.

Given the user-emitted wave/HSF contact, the operation of a PWE server
follows the steps shown in Fig.~\ref{fig:setup}. First, the HSF
impinging wavefront is detected, e.g., by using the HSF-synergistic
techniques of \cite{absense.2019,Liaskos.2019b}, or any other external
sensory system. Second, this information is relayed in a data packet
format to the PWE server via standard networking (e.g., a wired Ethernet
link). Third, the information is enqueued at the PWE server, along
with the sensory information for all HSF units in the environment,
and, fourth, is processed in due time (subject to scheduling). This
processing produces the intended configuration of the HSF, via a codebook-based
approach~\cite{Liaskos.b}. At the fifth step, the sensory information-matching
HSF configuration produced is relayed in data packet format to the
appropriate HSF unit via standard networking. Finally, at step six,
the HSF unit receives this data and changes it state accordingly.
The final step may optionally incorporate a HSF configuration hill-climbing
sub-step, based on a standard user device-to-BS control loop.

The total duration of this cycle can be expressed as:
\begin{equation}
\tau_{tot}=\tau_{s}+\tau_{n}^{\to}+\tau_{q}+\tau_{p}+\tau_{n}^{\gets}+\tau_{c},\label{eq:Ttot}
\end{equation}
where $\tau_{s}$, $\tau_{n}^{\to}$, $\tau_{q}$, $\tau_{p}$, $\tau_{s}^{\gets}$
and $\tau_{c}$ are the delays attributed to the six steps described
above. $\tau_{s}$ can be considered static but technology-dependent
(e.g., $10\mu\text{sec}$ for~\cite{Liaskos.2019b}, but $10\text{msec}$
for~\cite{absense.2019}). $\tau_{n}^{\to}$ and $\tau_{n}^{\gets}$
are generally subject to the networking delay, but may be considered
static and equal under the assumption of dedicated and symmetric network
paths. $\tau_{q}$ is subject to the scheduling policy of the PWE
server, as well as the total sensory load of the PWE network. $\tau_{p}$
can be static and even negligible in case of codebook-based HSF configuration
(e.g., $<1\mu\text{sec}$ or generally equal to a database lookup
query).

In general, the user will have moved by the end of the described cycle.
His dislocation, $d_{x}$, is a function of $\tau_{tot}$ and his
mobility pattern, $\overrightarrow{v}$:
\begin{equation}
d_{x}=f\left(\tau_{tot},\,\overrightarrow{v}\right).\label{eq:dx}
\end{equation}
We proceed to formulate the possible degradation in the power received
by the base station (user uplink) as a function of $d_{x}$. We employ
the notation of Fig.~\ref{fig:notation} and consider that $d_{x}=0$
at the beginning of step 1 of the PWE server cycle, without loss of
generality. The HSF will remain statically configured for $d_{x}=0$.

Let $\overrightarrow{I}_{x,d_{x}}$ be the directed power impinging
at point $x$ of the HSF, when the user is dislocated by $d_{x}$,
as shown in Fig.~\ref{fig:notation}, expressed as $\overrightarrow{I}_{x,d_{x}}=P_{x,d_{x}}\cdot\widehat{u}_{x,d_{x}}$,
where:
\begin{multline}
P_{x,d_{x}}\propto\frac{G_{x,d_{x}}^{Tx}}{\left(x-d_{x}\right)^{2}+\left(H-h\right)^{2}}\cdot\widehat{u}_{x,d_{x}}\,\\
\text{where}\,\widehat{u}_{x,d_{x}}=\frac{\left[\begin{array}{c}
x-d_{x}\\
H-h
\end{array}\right]}{\sqrt{\left(x-d_{x}\right)^{2}+\left(H-h\right)^{2}}},
\end{multline}
$G_{x,d_{x}}^{Tx}$ being the Tx antenna gain at $x$. The HSF configuration
is generally expressed as a tuple of impinging wave manipulators~\cite{Liaskos.TNET.2019},
$\left\langle \hat{n},\Delta\overrightarrow{J},\Delta\Phi,a\right\rangle \left(x\right)$,
where $\hat{n}\left(x\right)$ is a virtually rotated normal at point
$x$ of the HSF, defining the reflection of $\overrightarrow{I}_{x,d_{x}}$
as~\cite{Liaskos.TNET.2019}:
\begin{equation}
\overrightarrow{r}_{x,d_{x}}=\overrightarrow{I}_{x,d_{x}}-2\left(\overrightarrow{I}_{x,d_{x}}\cdot\hat{n}\left(x\right)\right)\hat{n}\left(x\right).
\end{equation}
Moreover, $\Delta\overrightarrow{J}\left(x\right)$ denotes a rotation
of the Jone's vector altering the polarization of $\overrightarrow{I}_{x,d_{x}}$,
while $\Delta\Phi\left(x\right)$ and $a\left(x\right)\in\left[0,1\right]$
quantify its phase and amplitude alterations by the HSF. We consider
an ideal HSF (i.e., with an infinite resolution of meta-atoms), where
each component of the HSF configuration tuple can be set and optimized
independently\footnote{In a practical HSFs with limited meta-atom numbers, $\hat{n},\Delta\overrightarrow{J},\Delta\Phi,a$
may be non-trivially co-joint at $x$, and also with tuple values
in an area around $x$. } and, subsequently focus only on $\hat{n}\left(x\right)$. For $d_{x}=0$,
redirecting $\overrightarrow{I}_{x,0}$ to the base station yields
the HSF configuration:
\begin{multline}
\hat{n_{*}}\left(x\right):\,\exists c>0:\\
\overrightarrow{I}_{x,0}-2\left(\overrightarrow{I}_{x,0}\cdot\hat{n_{*}}\left(x\right)\right)\hat{n_{*}}\left(x\right)=c\left[\begin{array}{c}
X-x\\
Y-\left(H-h\right)
\end{array}\right],\label{eq:5}
\end{multline}
i.e., a reflection that meets the base station location. The total
power reaching the base station can then be expressed as:
\begin{equation}
{\scriptscriptstyle P_{Rx}\left(d_{x}\right)=\stackrel[-x_{b}+d_{x}]{x_{b}+d_{x}}{\int}}C\left[\overrightarrow{I}_{x,d_{x}}-2\left(\overrightarrow{I}_{x,d_{x}}\cdot\hat{n}_{*}\left(x\right)\right)\hat{n}_{*}\left(x\right)\right]G_{x}^{Tx}dx,\label{eq:Pr}
\end{equation}
where $C\left[\overrightarrow{r}\right]=\left\Vert \overrightarrow{r}\right\Vert $
if $\overrightarrow{r}$ intersects a circle of radius $\delta$ positioned
at $X,Y$ (as shown in Fig.~\ref{fig:notation}), and $0$ otherwise.
The $x_{b}$ limits are employed to capture the main radiation span
of the Tx without loss of generality, and $G_{x}^{Tx}$ is the base
station antenna gain corresponding to $\overrightarrow{r}$. The $\delta$
quantity is introduced to capture a non-trivial antenna aperture size
at the distance between the reflection point and the base station.

Depending on $\hat{n}_{*}$, equation~(\ref{eq:Pr}) can yield power
degradation attributed to the user dislocation~$d_{x}$. In general,
this can be mitigated in three combine-able ways:
\begin{itemize}
\item Minimize the networking delay, $\tau_{n}^{\to}+\tau_{n}^{\gets}$,
and the queuing delay, $\tau_{q}$, via appropriate routing and scheduling
techniques, thereby minimizing the user dislocation via relation~(\ref{eq:dx}).
\item Forecast the expected user location at time $t+\tau_{tot}$ and produce
the corresponding HSF configuration at step $4$ (rather than configuring
it for time $t$).
\item Make the HSF configuration aware of the potential user dislocation
and mitigate its effects on the received power.
\end{itemize}
In the next Section we study the latter approach.

\section{The proposed, mobility-aware beam steering\label{sec:The-proposed,-mobility-aware}}

The key-idea behind the proposed beam-steering approach is to exploit
the precise wave control offered by the metasurfaces, to cover the
full span of the possible Tx positions. This can be accomplished by
tuning the HSF configuration, $\hat{n}\left(x\right)$, to reflect
impinging waves from any probable user location (Tx) to the base station
antenna (Rx).

\begin{figure}
\begin{centering}
\textsf{\includegraphics[width=0.9\columnwidth]{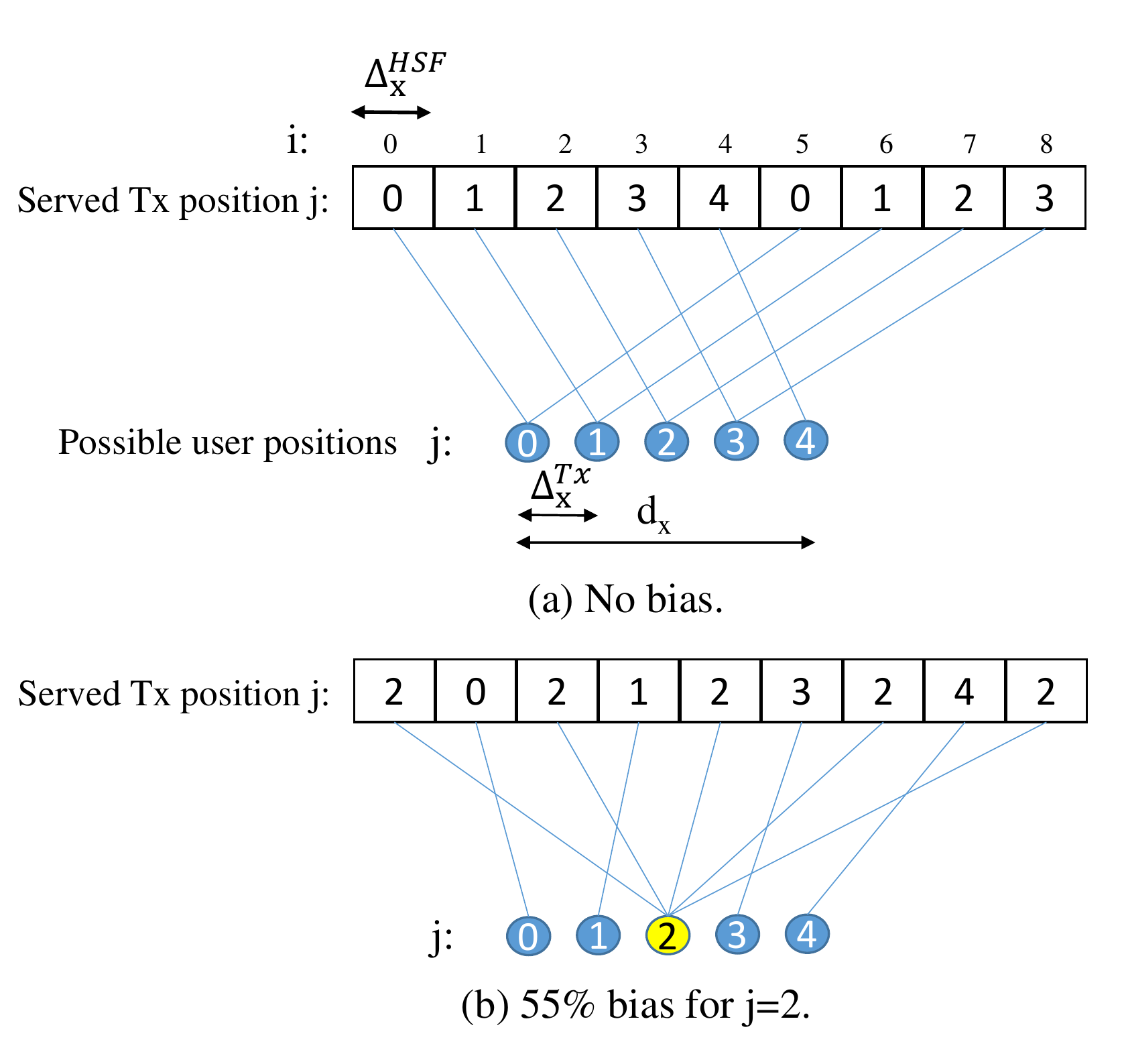}}
\par\end{centering}
\caption{\label{fig:proposed}Visualization of the proposed beam steering approach
(a) without bias, and (b) with bias (i.e., a degree of certainty)
for a possible user location.}
\end{figure}
For ease of exposition, we discretize the HSF into sub-units of length
$\Delta_{x}^{HSF}$, over which the $\hat{n}$ configuration remains
constant, as shown in Fig.~\ref{fig:proposed}. For a HSF of total
length $L$, we index each sub-unit as:
\begin{equation}
i=0,\ldots I,\,I=\left\lceil \frac{L}{\Delta_{x}^{HSF}}\right\rceil .
\end{equation}
In a similar manner, we discretize the possible user locations with
a step of $\Delta_{x}^{Tx}$, and index them as:
\begin{equation}
j=0,\ldots J,\,J=\left\lceil \frac{d_{x}}{\Delta_{x}^{Tx}}\right\rceil .
\end{equation}
Subsequently, let $\hat{n}_{ij}$ be defined as the configuration
deployed at sub-unit $i$ of the HSF, steering the reflection from
user position~$j$ to the base station, i.e., similarly to relation~(\ref{eq:5}):
\begin{multline}
\hat{n}_{ij}:\,\exists c>0:\overrightarrow{I}_{i\cdot\Delta_{x}^{HSF},j\cdot\Delta_{x}^{Tx}}-2{\scriptstyle \left(\overrightarrow{I}_{i\cdot\Delta_{x}^{HSF},j\cdot\Delta_{x}^{Tx}}\cdot\hat{n}_{ij}\right)}\hat{n}_{ij}=\\
=c\left[\begin{array}{c}
X-x\\
Y-\left(H-h\right)
\end{array}\right].
\end{multline}
Based on the $\hat{n}_{ij}$ definition, we proceed to define a \emph{user
position-unbiased HSF configuration beam steering}, illustrated in
Fig.~\ref{fig:proposed} (top), as follows:
\begin{equation}
\hat{n}_{ij}:\,j=mod\left(i,J+1\right),\,i=0\ldots I.\label{eq:unbiased}
\end{equation}
The unbiased configuration ensures that at least a portion of the
HSF-impinging power will reach the base station. It is intended as
a connectivity-ensuring approach, when no further information exists
about the location of the user.

We proceed to study the case whether exists an increased confidence
for the correct user position, $j_{c}$, e.g., expressed as a probability,
$p_{j_{c}}$, with regard to the other possible positions. In this
case, the proposed beam steering becomes \emph{biased} to the outstanding
user position. The process is illustrated in Fig.~\ref{fig:proposed}
(bottom). First, we set:
\begin{equation}
\hat{n}_{ij_{c}}:\text{for\,}i=n\cdot\Delta i,\,\Delta i=\left\llbracket p\cdot\left(I+1\right)\right\rrbracket ,n=0,1,2,\ldots\label{eq:biased}
\end{equation}
Then, all $i$ values not covered by relation~(\ref{eq:biased})
are set to a $j$ value that follows a round-robin cycle over all
possible j values except for $j_{c}$.

It is noted that in the biased case, all user positions apart from
the outstanding one are inherently assumed to share uniformly the
probability of $1-p$. Producing a configuration that takes into account
any probability distribution position over $j$ is a directly applicable
generalization, by using techniques from the field of periodic scheduling
with multiple concurrent weights~\cite{Liaskos.2012f}.

\section{Evaluation\label{sec:Evaluation}}

We evaluate the proposed beam steering approach via ray-tracing based
simulations. The employed tool has been presented in~\cite{Liaskos.TNET.2019}.

The simulations replicate the setup of Fig.~\ref{fig:setup}, i.e,
an open-ended corridor of total height $H=3$m and total length $L=5$m.
A user device moves at height $h=1$m, starting with an offset of
$o=1$m from the left side of the corridor. The coordinate system
origin is set at this point. The base station is located at $X=3.6$m,
$Y=1.4$m. The user antenna has a beam angle of $30^{o}$ with uniform,
unitary gain and is pointed upwards and in parallel to the y-axis.
The base station antenna is similar, but with a beam angle of $60^{o}$
and oriented towards a counter-clockwise angle of $77^{o}$ with regard
to the y-axis. The system operates at $60$~GHz, with the user device
(Tx) emitting at a power level of $20$~dBm and the base station
acting as the receiver (Rx).

The HSF-coated ceiling seeks to maximize the $\nicefrac{P_{rx}}{P_{tx}}$
ratio, i.e., ideally directing the totality of the user's emission
to the base station. In order to emulate a lossy metasurface, we consider
a resolution of $\Delta_{x}^{HSF}=1$mm across the HSF. In other words,
a configured virtual surface normal value is necessarily constant
and uniform for $\Delta x$. This assumption yields a metasurface
of medium efficiency, given that for the studied wavelength $\lambda=5$mm,
highly efficient metasurface functionality is typically achieved for
control at a resolution level of less than $\nicefrac{\lambda}{10}=0.5$mm~\cite{MSSurveyAllFunctionsAndTypes}.
The floor is treated as a perfect conductor and $\Delta_{x}^{Tx}=2\text{mm}$.

The HSF is configured correctly for maximizing the $\nicefrac{P_{rx}}{P_{tx}}$
ratio only when the user is located at the coordinate system origin.
We are interested in the degradation of the $\nicefrac{P_{rx}}{P_{tx}}$
ratio as the user is displaced towards the x-axis while the HSF configuration
remains the same. The regular (non-HSF) propagation is compared to
the HSF-enabled propagation under several parameterizations (bias
values) of the proposed beam steering.

\begin{figure}
\begin{centering}
\textsf{\includegraphics[width=1\columnwidth]{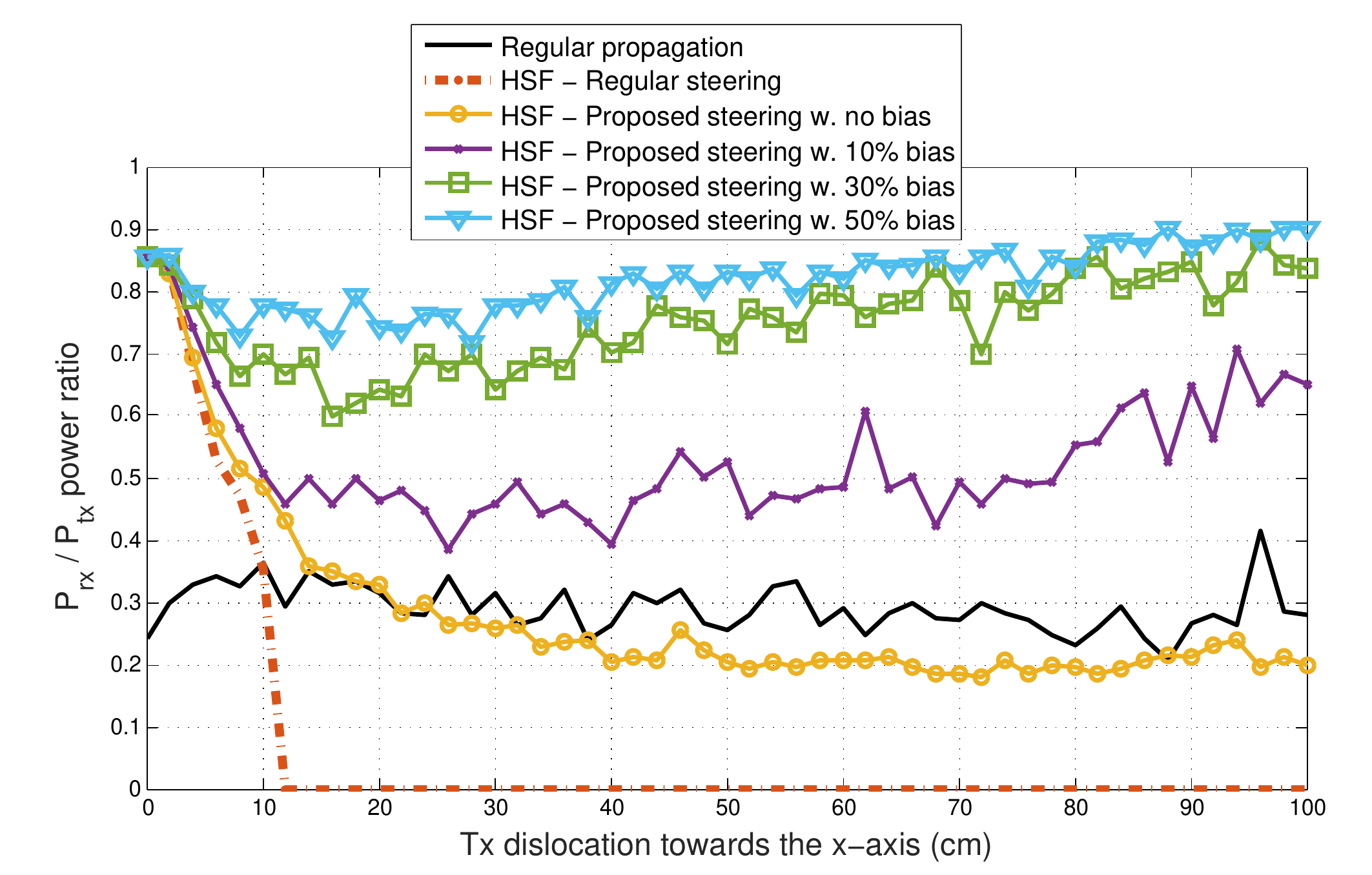}}
\par\end{centering}
\caption{\label{fig:results}$\nicefrac{P_{rx}}{P_{tx}}$ ratio of various
propagation and beam steering approaches, as the user is displaced
towards the x-axis, under static HSF configuration. Notice that the
dislocation brings the user physically closer to the base station.}
\end{figure}
\begin{figure}
\begin{centering}
\textsf{\includegraphics[width=1\columnwidth]{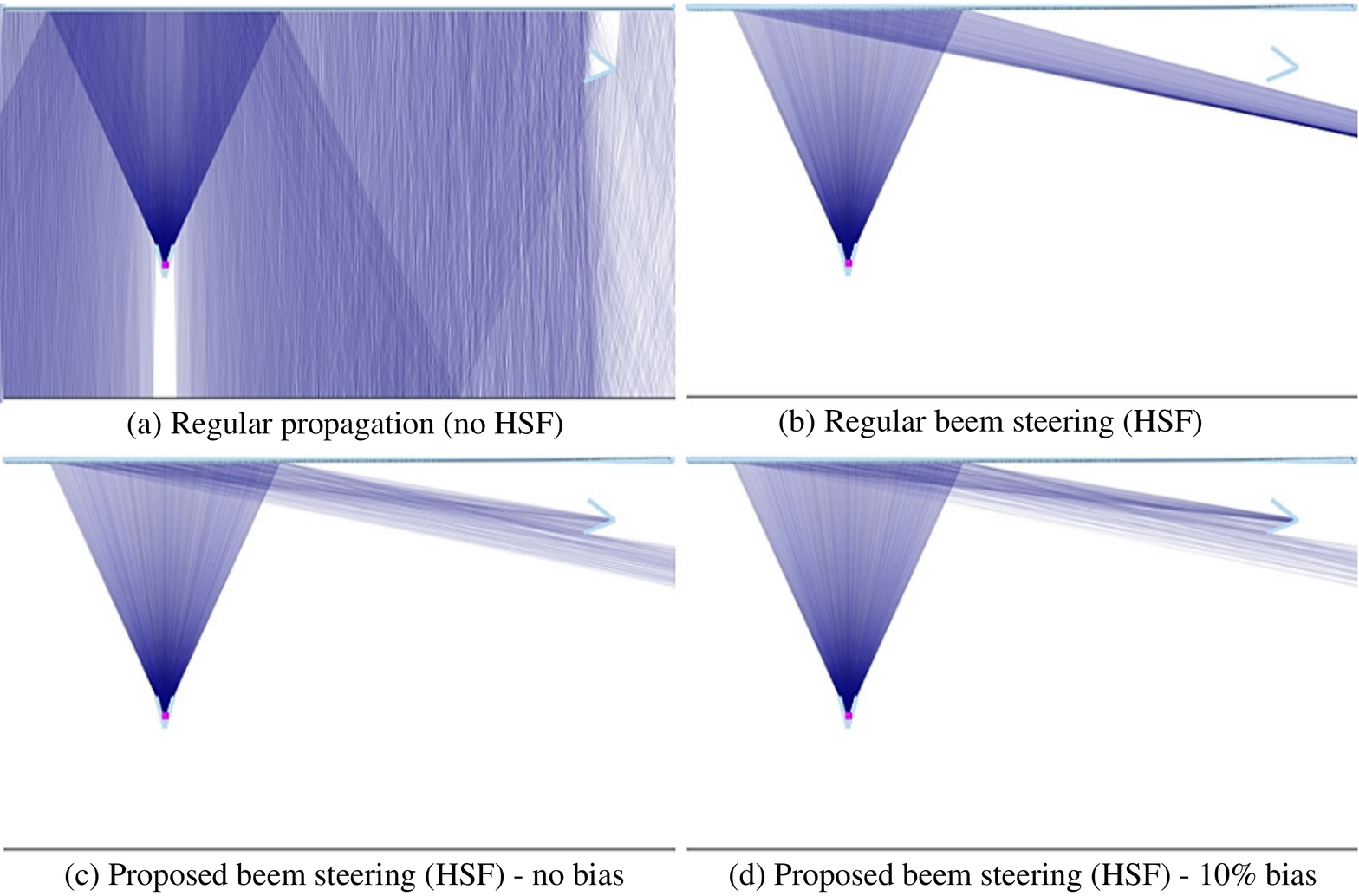}}
\par\end{centering}
\caption{\label{fig:resultsRT}Ray-tracing based illustrations of the propagation
at $20$cm Tx dislocation.}
\vspace{-10bp}
\end{figure}

The results are shown in Fig.~\ref{fig:results}, with indicative
ray-tracing results given in Fig.~\ref{fig:resultsRT}. First, the
regular propagation (non-HSF assisted) case yields a $\nicefrac{P_{rx}}{P_{tx}}$
ratio of approximately $30$\%, given the asymmetry of the Tx-Rx setup
and the lack of any natural mechanism for focusing the propagating
rays. The efficiency remains almost static during the Tx displacement
(noting that this movement reduces the geometric distance of the Tx-Rx
pair). All HSF-assisted cases yield an efficiency of $86$\% when
there is no misalignment (Tx dislocation is $0$). The $14$\% loss
is due to metasurface losses, as described above.

Second, the HSF-assisted case that employs regular, static beam steering
quickly exhibits a drop in its efficiency, which becomes null (i.e.,
worse than the regular propagation) at a displacement of just $11$cm.
While such a displacement is significant in mm-wave communications
in general, the nullification of the efficiency showcases the necessity
for novel beam steering approaches such as the proposed one. Indeed,
even with no Tx position bias, the proposed beam steering retains
the maximal efficiency of $86$\% while ensuring near-regular propagation
performance in the worst case.

Third, in the case of steering with bias (i.e., guessing the location
of the Tx with a very small degree of certainty), even a $10$\% certainty
yields robust performance improvement for any Tx dislocation, which
is also significantly better than the regular propagation in any case.
At $30$\% bias, the efficiency is nearly maximized, as exhibited
by the fact that subsequently increasing the bias to $50$\% yields
disproportional gains in efficiency. These modest certainty levels
could be potentially attained by the vast array of mobility predictors
that exist in the literature~\cite{MobilityPrediction.2018d,Zhang.2018m}.
Evaluating the synergy between existing predictions and the proposed
mobility-aware beam steering algorithm constitutes an open research
challenge.

\section{Conclusion and Outlook\label{sec:Conclusion}}

Software-defined metasurfaces enable precise control over the wireless
propagation phenomenon, tailoring it to the needs of users, mitigating
distance, fading and mobility effects on wireless communications.
The paper studied the effects of misalignment between metasurface
configurations and the wireless emissions of the users, which can
appear due to latency in the sense/reconfigure cycle of programmable
wireless environments. An analytical model was proposed to quantify
the misalignment effects, followed by a novel beam steering approach
that can mitigate potential misalignment by exploiting the wave manipulation
precision offered by metasurfaces.

Further directions can extend the proposed analytical model to a full
3D scenario, while taking into account the complete spectrum of wavefront
shaping capabilities offered by metasurfaces. Moreover, the proposed
system model can constitute the basis for analyzing misalignment problems
from a queuing theory perspective, studying the scheduling problem
of sensing and configuring a programmable wireless environment for
multiple users. In that aspect, the Age of Information concept can
constitute a promising methodological approach~\cite{Kadota.2018AoIInfocomBestpaper,pappasAoI.2019}.

\section*{Acknowledgment}

This work was funded by the European Union via the Horizon 2020: Future
Emerging Topics call (FETOPEN), grant EU736876, project VISORSURF
(http://www.visorsurf.eu).


\end{document}